# A first-principle calculation of the XANES spectrum of $Cu^{2+}$ in water

G. La Penna[1,a)], V. Minicozzi[2,3], S. Morante[2,3,b)], G.C. Rossi[2,3,4] and F.Stellato[2,3,c)]

[1] *CNR - Institute for Chemistry of Organometallic Compounds, Sesto Fiorentino, 50019, Italy*

[2] *INFN, Rome "Tor Vergata" , 00133, Italy*

[3] *Department of Physics University of Rome "Tor Vergata" , 00133, Italy*

[4] *Centro Studi e Ricerche "Enrico Fermi", Roma, 00184, Italy*

The progress in high performance computing we are witnessing today offers the possibility of accurate electron density calculations of systems in realistic physico-chemical conditions. In this paper, we present a strategy aimed at performing a first-principle computation of the low energy part of the X-ray Absorption Spectroscopy (XAS) spectrum based on the density functional theory calculation of the electronic potential.

To test its effectiveness we apply the method to the computation of the X-ray Absorption Near Edge Structure part of the XAS spectrum in the paradigmatic, but simple case of $Cu^{2+}$ in water. In order to keep into account the effect of the metal site structure fluctuations in determining the experimental signal, the theoretical spectrum is evaluated as the average over the computed spectra of a statistically significant number of simulated metal site configurations. The comparison of experimental data with theoretical calculations suggests that $Cu^{2+}$ lives preferentially in a square-pyramidal geometry.

The remarkable success of this approach in the interpretation of XAS data makes us optimistic about the possibility of extending the computational strategy we have outlined to the more interesting case of molecules of biological relevance bound to transition metal ions.

[a)] Associated to *INFN, Rome "Tor Vergata"*
[b)] Associated to *CNR - Institute for Chemistry of Organometallic Compounds, Sesto Fiorentino*
[c)] Author to whom correspondence should be addressed. stellato@roma2.infn.it

## I. INTRODUCTION

X-ray absorption spectroscopy (XAS) has become an increasingly important tool for probing the local environment around selected atoms both in solids and liquids and it is therefore an ideal tool for structural investigations of non-crystalline samples. For this reason, it has been widely applied to the study of the atomic arrangement around metal ions in complex with proteins and peptides. Application of this spectroscopic technique has proved to be especially useful in the study of the coordination site in the very relevant case of systems formed by metal ions (mostly Zn and Cu) bound to amyloid-β (Aβ) peptides [1-3] or to the prion protein [4-7]. Because of their enormous biomedical relevance, compounds of this kind have also been extensively studied *in silico* by classical and *ab initio* molecular dynamics (MD) simulations [8-11].

It would then be extremely important to extend the realm of application of theoretical methods to arrive at a first-principle computation of the XAS spectrum of a metal-peptide complex. This step would represent the necessary bridge between *in silico* and *in vitro* experiments, allowing on the one side to provide a reliably supported interpretation of experimental data, and on the other an experimental validation of computational results. As already indicated above several groups, including ours, have worked in this direction in recent years using a number of approaches, focusing on different computational methods and investigating different regions of the XAS spectrum.

From the experimental point of view, a XAS spectrum is commonly divided into two energy regions [see e.g. [12]], XANES (X-ray Absorption Near Edge Structure) and EXAFS (Extended X-ray Absorption Fine Structure). The XANES region, which starts a few eV before the edge energy[a] and extends some few tenths of eV above it, is dominated by a variety of very complicated low energy transfer events.

This part of the spectrum contains detailed information about the local atomic structure around the absorber, but its theoretical analysis is quite complicated and requires some preliminary knowledge of the geometrical arrangement of the atoms surrounding the absorber and a very good model for the interatomic potential seen by the outgoing photo-electron. Conventionally the EXAFS region starts where the XANES region ends and extends a few hundreds eV above the edge. In the EXAFS region, the photo-electron energy is substantially larger than in the XANES one, so that single scattering events are dominant. Multiple scattering contributions are still present and must be taken into account if a refined analysis of experimental data is required as in the case of biological samples. The quantitative analysis of the EXAFS region is less difficult than the XANES one, but the XANES region potentially contains a lot more structural information about the metal binding site that would be of the utmost importance to unveil.

---

[a] The N-th edge, with N = K, L, M, …, is the minimal energy necessary to extract an electron from the N-th shell in the continuum.

Computationally, both classical and *ab initio* MD have been used to produce sets of geometrical configurations that are successively used either as a starting point for structural fits or directly to calculate the associated XAS spectra. Various more or less sophisticated theoretical approaches have been employed by different groups [13-17] in the especially challenging case of the analysis of the XANES region. Each combination of computational and simulation strategies has its pros and cons that make it suitable for addressing specific classes of problems.

For example, our group has used in the past an integrated first-principle simulation of the EXAFS spectrum of biologically interesting metal-protein compounds, like Aβ peptides and prion proteins complexed with $Cu^{2+}$ and/or $Zn^{2+}$ [10]. This approach, owing to the detailed atomic information that numerical methods can provide, has allowed a convincing interpretation of experimental XAS data, thus helping in successfully discriminating among different structural models.

Roscioni *et al.* [13] have proposed a general strategy for the computation of the XANES part of the XAS spectrum of compounds in complex with metal ions that, as a first application, was employed in the case of $Ni^{2+}$ ions in water. The idea is to produce individual configurations of the system via classical MD simulations and compute the XANES spectrum of each configuration exploiting the basic approach implemented in the MXAN code [18]. Without resorting to any optimization in the space of structural parameters, the theoretical spectrum is finally computed by averaging over the spectra evaluated on each one of the MD generated configurations. The comparison of experimental data and theoretical calculations turns out to be quite satisfactory.

More recently, D'Angelo *et al.* [14] have successfully extended this approach to the calculation of the XANES spectrum of the Fe-protein neuroglobin. works suggest that a strategy in which the XANES spectrum is evaluated as an average over the spectra of a number of selected configurations (chosen along different MD trajectories) is able to effectively take into account also the static disorder effect of non-crystalline and glassy samples and is therefore appropriate to perform a quantitative analysis of the spectra of disordered systems.

We stress that both works make use of the MXAN code in which the XANES photo-absorption cross-section is computed by fully including multiple scattering contributions, though within the crude but largely employed muffin-tin approximation for the photo-electron[b].

Other attempts have been developed in the literature aiming at a reliable first-principle calculation of the XANES spectrum. Among them it is worth mentioning the seminal paper of Rehr and co-workers [15] where the largely employed FEFF code was introduced. In this work the XAS spectrum is computed relying on the real-space Green's function (RSGF) approach within the quasi-particle picture. This choice allows a more accurate treatment than in other earlier approaches of several many-body effects. A detailed justification of the RSGF approach is provided by the quasi-boson theory [20] where it is shown that the RSGF can be viewed as the effective propagator in the presence of a core hole with multi-electron effects treated in terms of the appropriate spectral function.

In the present paper we illustrate and discuss a general strategy aimed at performing a parameter-free first-principle calculation of the XANES spectrum in the case of systems in solution, combining the *ab-initio* computational strategy of Gougoussis *et al.* [16,17] for the evaluation of the electronic density with the idea of averaging over separately computed spectra to take into account the disorder effects of non-crystallized samples.

In this work, the difficulties inherent in dealing with density-functional theory (DFT) calculations with transition metal atoms (huge kinetic-energy cutoffs and large supercells with reduced symmetry) are largely solved by employing ultrasoft pseudopotentials (rather than norm-conserving ones) in the valence electron wave function construction. Without spoiling the possibility of using the Lanczos recursion method of continued fractions [21] for the calculation of the absorption cross section (thanks to an appropriate reformulation of the approach [16,17]), the use of ultrasoft pseudopotentials has also the non-negligible benefit of reducing the computational cost by almost one order of magnitude. Furthermore the *ab initio* quantum-mechanical calculation of the electronic charge density (in the presence of a core-hole) has also the great merit of allowing the evaluation of the XANES spectrum without the need of relying on any approximate form of the potential like the widely used muffin-tin expression.

---

[b] The muffin-tin approximation consists in assuming a spherical scattering potential centered on each atom and a constant value in the interstitial region between atoms [19].

As for the system configurations whose XANES spectrum is computed and compared with experimental data, the proposal advocated in the present paper is to employ quantum-mechanical optimized configurations taken along some classical MD simulation trajectory. Naturally the method could be straightforwardly extended to get the input configurations from entirely quantum-mechanical (e.g. Car-Parrinello [22]) MD simulations if desired. Indeed, the idea of combining molecular dynamics and *ab initio* XAS calculations has already been exploited by Galli and co-workers [23,24] to study the structure of water in different conditions and in the presence of different ions.

In the present case, we apply a similar approach to the study of a transition metal K-edge spectrum, making use of the method described in ref. [16,17]. The novelty of our work is to propose a simple protocol to afford the simulation of XAS spectra when metal ions are bound to disordered ligands and where in the process of sampling ligand conformations structural information coming from experiments needs to be injected.

Before daring to move to more complicated (and biologically more interesting) systems, we present here, as a test case, a thorough study of the application of this strategy to the calculation of the XANES spectrum of the apparently simple case of $Cu^{2+}$ in solution.

Previous studies agree on the stable presence of four equatorial water ligands, but there are difficulties in detecting the number and even the presence of axial water ligands [25].

In the last few years several works have been published in which quantum mechanical calculations are performed to elucidate the structure of $Cu^{2+}$ ions in aqueous solvents [26,27,28] and in the gas phase [28]. Interestingly in these works it has been proposed that clusters with different coordination numbers may coexist in solution [29]. Efforts have also been devoted to computing the ultraviolet absorption spectrum of $Cu^{2+}$ in aqueous solution with help of DFT based molecular dynamics [30]. Recent [31] and less recent works [32,33] have been devoted to the EXAFS and XANES analysis of solvated $Cu^{2+}$ complexes, suggesting that the dominant coordination mode is a square based pyramidal geometry. Unlike what is done in these works, where the emphasis was on finding the best agreement between calculated and experimental spectral data by an overall structural fitting-based procedure, our effort in the present paper is focused on setting up a method capable of yielding a fully *ab initio* calculation of the XANES part of the X-ray spectrum without any optimization of structural parameters. In this way we only rely on a first-principle calculation of the electronic density and consequently of the associated potential seen by the photo-electron. We consider this last ingredient a major computational improvement over the use of more crude approximation for the potential like the one provided by formulae of the muffin-tin type.

The theoretical setting we propose can thus yield a more fundamental framework for the simulation of the XANES spectra of disordered systems that would significantly reduce the unavoidable bias associated with modeling the atomic structure around the absorbing center.

In this paper we present a fully integrated investigation, where the theoretical approach we are advocating is successfully tested against the experimental XANES data of $Cu^{2+}$ in water we acquired at the European Synchrotron Radiation Facility (ESRF - Grenoble).

## II. MATERIALS AND METHODS

### A. X-ray data acquisition

A 10 mM $Cu^{2+}$ solution was prepared by dissolving $CuSO_4$ in millipore water. XAS experiments were carried out at the BM30B beamline at the European Synchrotron Radiation Facility (Grenoble, France) [34]. The beam energy was selected using a Si(220) double-crystal monochromator with a resolution of about 0.5 eV [35]. The beam spot on the sample was approximately $300 \times 200$ μm[34] ($H \times V$, FWHM). Spectra were recorded in fluorescence mode by the use of a 30-element solid-state Ge detector. To avoid photo-degradation and spectra evolution during XAS measurements, all the samples were first rapidly brought from room temperature to 77 K (liquid nitrogen) and then cooled and kept at 13 K in a liquid helium cryostat throughout all XAS measurements. Three scans, each one taken at a different position of the sample, have been acquired and averaged. The spectra were then normalized using the standard procedure implemented in the Athena package [36].

### B. Computational methods

Structural models of $Cu^{2+}$ in water, with water bound in different coordination modes, have been prepared by classical MD employing empirical force-fields according to the following procedure. A $Cu^{2+}$ ion was put in the middle of a 18.774 Å side box filled with 213 water molecules and including 2 $Cl^-$ ions for neutrality. The system was constructed by taking an equilibrated snapshot of 216 TIP3P water molecules [37] and replacing the O atom of one randomly chosen water molecule with a $Cu^{2+}$ ion and two other, equally randomly chosen water molecules, with two $Cl^-$ ions.

Exploiting the so-called cationic dummy atom method [38], positive charges were located in different points around the $Cu^{2+}$ ion, all lying at a 0.9 Å distance from the $Cu^{2+}$ center.

The structure and the charge of the dummy atoms was step-wise modified every 1 ns, starting from a system with no dummy atoms (non-coordinating model) and moving to a system with two 1/3 (in $e^-$ units) point charges (digonal coordination model), then with four 1/3 charges on the vertices forming the square base of the octahedron (square-planar coordination model, $M^0_{sqp}$) and finally with six 1/3 charges on each of the 6 vertices of the octahedron (octahedral coordination, $M^0_{oct}$). The negative chloride charge is modified accordingly, so as to have always zero total charge in the simulation box. The inclusion of explicit counterions in the simulation cell in classical MD is a routine strategy when dealing with electrostatic interactions [39].

The final dummy ion model was taken to have an octahedral symmetry, with a single Lennard-Jones (LJ) site in the center and point-like charges displaced at the six vertices. The $\sigma$ and $\varepsilon$ values of the LJ central site were taken to be 3.1429 Å and 2.77 cal/mol, respectively. Arithmetic combinations of the $\sigma_{LJ}$ parameters were formed to determine the interaction potential between different LJ sites. The TIP3P water model [37] and the usual chloride model [40] were adopted.

The time-step of the simulation was 0.5 fs. The long-range electrostatic interactions were approximated with the smooth particle-mesh Ewald (PME) method [41], using a cut-off of 9 Å and a mesh with a grid spacing of 0.1 Å. The PME direct space tolerance was set to $10^{-4}$ kcal/mol. Classical MD simulations were carried out using the NAMD code [42] at the temperature of 300 K using a stochastic thermostat [43].

For each cation model, i.e. for each of the four arrangements of dummy charges around the ion described above, a 1 ns long MD trajectory was produced and 20 configurations, one every 50 ps, were extracted and stored for the further analysis.

The radial distribution function (RDF) of the O belonging to water molecules around the cation center was computed and monitored all along the MD trajectory. Besides the first RDF peak, by increasing the cation charge, a second peak becomes appreciable. For the highest value of cationic charge (2 = 6 x 1/3 charges on the octahedron vertices - $M^0_{oct}$ model) first and second peak together include, on average, 29 water molecules. According to this fact, we decided to consider for the further quantum-mechanical (QM) calculation of the potential seen by the photo-electron a system with 29 water molecules around the Cu center, thus including also the second Cu solvation shell.

Chloride atoms fell within this sphere only in one tenth of the configurations and were never found within the first $Cu^{2+}$ coordination shell. In the following for short we will sometimes refer to this system as the "2$^{nd}$ sphere model".

The collected configurations, even those with 4 and 6 positive charges, are supposed to represent approximated descriptions of *aquo* Cu ions. To improve the quality of this approximation we performed a partial geometry optimization (or energy relaxation) of the 2$^{nd}$ sphere model of each stored configuration.

For the QM calculation of the potential a DFT approach, widely used for large atomic systems (from 100 to 1000 atoms), was employed. The QuantumESPRESSO (v.5.0.2) package [44] was adopted because of its efficient implementation of ultrasoft pseudopotentials which makes affordable core-valence interactions calculations within plane-wave electronic structural representation. The spin state of the system was taken to always be a doublet ($S_z = 1$, one single unpaired electron) with all the Kohn-Sham states, except the highest in energy, occupied by 2 electrons.

Each one of these system configurations (we recall, formed by 29 water molecules and one $Cu^{2+}$ ion) was inserted in a cubic super-cell with a side of 2.2 nm with the $Cu^{2+}$ ion at its center. The empty space within the 29 explicit water molecules and the cell sides was filled with a uniform dielectric mimicking the bulk liquid water at room conditions [45]. A smooth kinetic energy cut-off of 30 Ry was used, together with a finer charge density cut-off of 300 Ry.

The resulting structures were relaxed by minimizing the total energy via the Broyden-Fletcher-Goldfarb-Shanno quasi-Newton minimization algorithm. In all cases minimization is stopped after 30 steps. The Martyna-Tuckerman correction [46,47] to both total energy and electron density was introduced to account for the non-zero charge distribution within the periodic super-cell.

Finally, the XANES spectrum for each selected and relaxed configuration was computed (in the dipole approximation) with the help of the XSpectra [16,17,48] code of the QuantumESPRESSO suite. XSpectra is a post-processing tool that relies on the output electronic charge density evaluated with the help of the PWscf code of QuantumESPRESSO. To simulate core-hole effects a pseudopotential with a hole in *s* state is used to describe the Cu electronic state after X-ray absorption. The Lorentzian broadening parameter used for the XANES calculation is 2.0 eV. A schematic flow-chart of our calculation strategy is displayed in Figure 1.

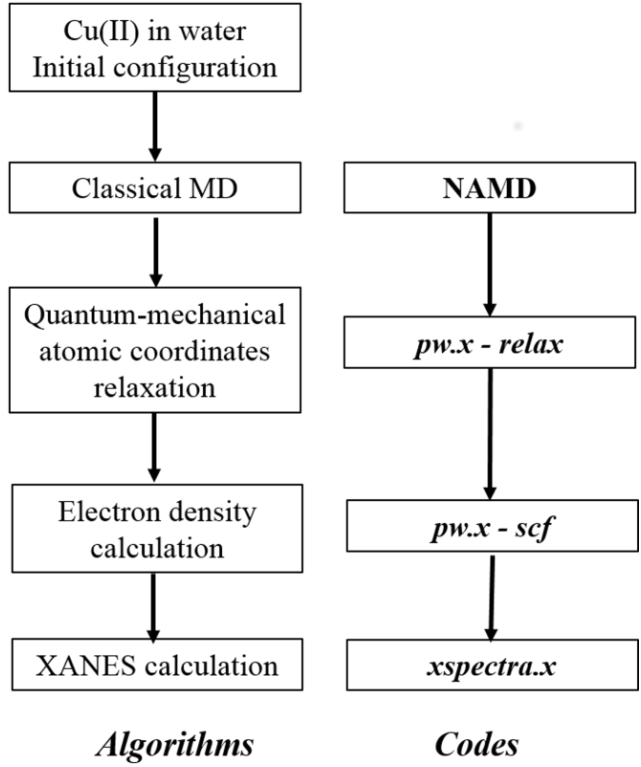

**FIG. 1.** Flow-chart summarizing our computational strategy. The algorithms used for the calculation are indicated on the left and the codes used for each step on the right.

## III. RESULTS AND DISCUSSION

As we said, in the $M^0_{sqp}$ model $Cu^{2+}$ is initially tetra-coordinated to water molecules in a square planar geometry, while in the model $M^0_{oct}$ $Cu^{2+}$ is hexa-coordinated to water molecules in an octahedral geometry. In Figure 2 we show typical representative structures of the systems obtained from classical MD simulations, before (panel (a) and (b), structures $M^0_{sqp}$ and $M^0_{oct}$) and after the QM relaxation step (panel (c) and (d)). We call $M^{rel}_{sqp}$ and $M^{rel}_{oct}$ respectively the structures obtained after relaxation.

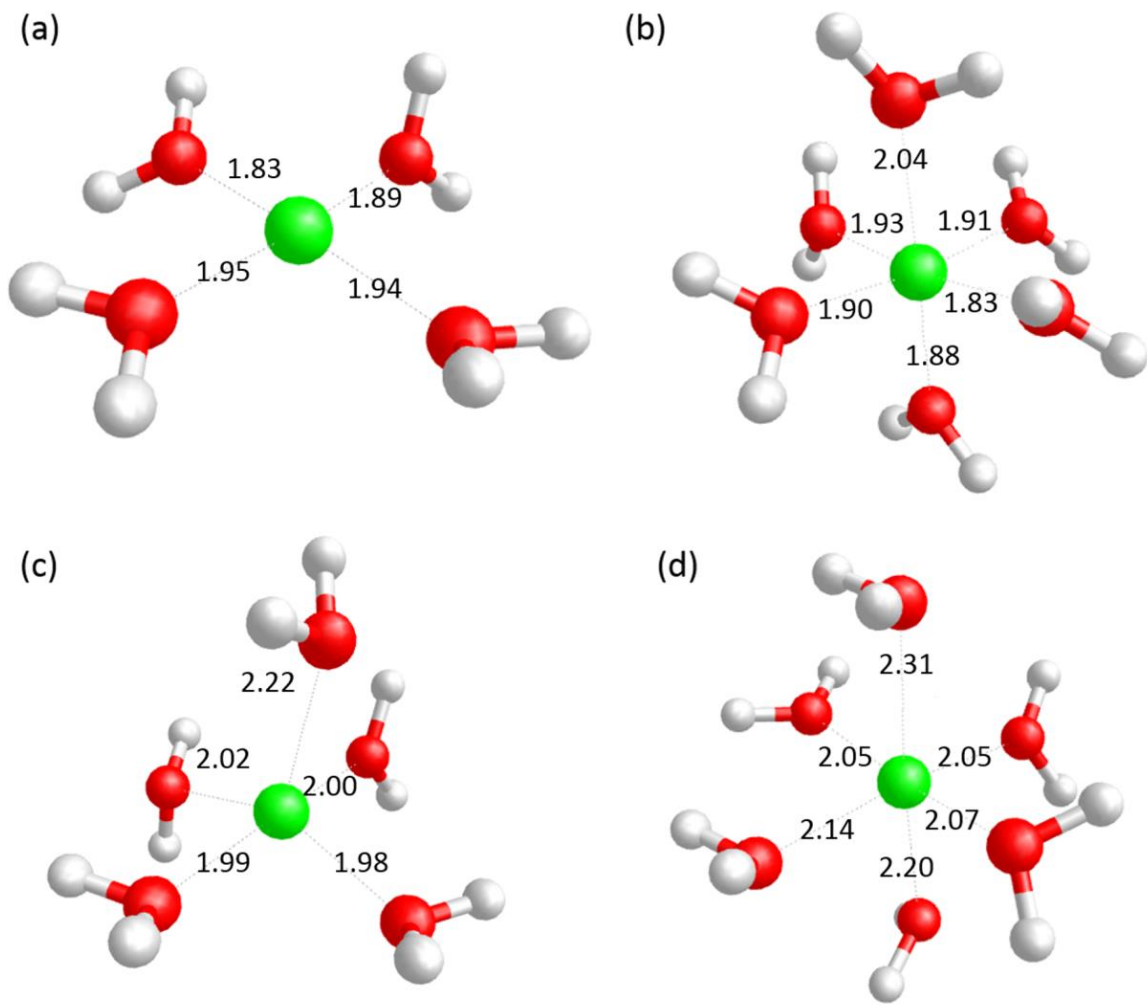

**FIG. 2.** $Cu^{2+}$ coordination modes in water from classical MD simulations. Representative square planar (*sqp*) and octahedral (*oct*) geometries before nuclei position relaxation (panel (a) and panel (b)) and evolution of the same configuration after nuclei position relaxation (panels (c) and panel (d)) are sketched. Only water molecules having oxygen atoms within a sphere of radius $r = 2.50$ Å from $Cu^{2+}$ are displayed. Distances are in Ångström. Cu is green, hydrogen is white and oxygen is red.

**Table I**. The range of variation of the root-mean-square displacements (RMSD) between all the pairs of the two sets of selected configurations before and after relaxation is calculated with the VMD program [49]. We report data including either all the 29 oxygen atoms of the water molecules, excluding H atoms ($RMSD_1$), or of only the water oxygen that lie within a sphere of $r = 2.5$ Å from Cu ($RMSD_2$). In the fourth column we give the metal site coordination number and in the last column the bond valence sum [50] (BVS) value relative to the coordinated atoms around the Cu center.

Despite the fact that in principle complete QM energy relaxation processes, even if starting from

| Sample | RMSD$_1$ (Å) (all atoms, excluding H) | RMSD$_2$ (Å) (atoms within 2.5 Å, excluding H) | Coordination number | BVS |
|---|---|---|---|---|
| Before relaxation | | | | |
| $M_{oct}^0$ | 5.2 ± 0.3 | 1.6±0.4 | 6 | 2.7 ÷ 3.2 |
| $M_{sqp}^0$ | 5.3 ± 0.3 | 1.1+0.7 | 4 | 1.9 ÷ 2.4 |
| After relaxation | | | | |
| $M_{oct}^{rel}$ | 5.2 ± 0.3 | 1.6±0.4 | 6 | 1.7 ÷ 1.9 |
| $M_{sqp}^{rel}$ | 5.2 ± 0.3 | 1.5 ±0.6 | 5 | 1.7 ÷ 2.1 |

possibly different initial second solvation shell structures, are expected to end in the same global minimum, constraints and local minima trapping forbid reaching this unique final state.

Most of the decrease in energy in the QM minimization is due to relaxation of bond distances towards the equilibrium values, while non-bonding interactions assist these local changes. Fast changes of coordination number and geometry around the cation, occurring within the 30 minimization steps we decided to take, reflect these two components of the relaxation process. On a longer time-scale other processes would occur. For instance, at a later stage we expect the interactions between the relaxed first coordination sphere and the surrounding water molecules to slowly settle down.

The idea of allowing for a "short" relaxation (30 steps of energy minimization, see Methods), without reaching a complete relaxation of the entire coordination sphere (we recall, made by 29 water molecules in the mean-field approximation of the bulk water) is meant to mimic the events occurring in the fast cooling to which the sample was subjected in the actual XAS experiments. In any case we must observe that in any model of interatomic forces (both empirical and at the DFT level of approximation), the global minimum is never too well separated in energy from other configurations in a system with a large number of degrees of freedom.

The $Cu^{2+}$ ion has a single vacancy in the highest occupied KS state with a strong *d* character (data not shown here) and contributes to lower the screening of ligands along the coordination directions (the Jahn-Teller effect). Starting from the highly symmetric *sqp* and *oct* structures, the configurations that are attained when approaching energy minima exhibit significant distortions (see Fig. 2c and d, respectively) as a manifestation of the Jahn-Teller effect. Our analysis shows that, interestingly, comparison with the XAS experiments allows identifying the precise geometrical properties of the distortion.

In Table I we report the root-mean square displacements (RMSD) between pairs of configurations (see Methods for details) within each set of the selected configurations. The RMSD variability range computed including all the oxygen atoms (RMSD1) indicates the existence of significant deviations among structures, while the smaller value of the RMSD of Cu-coordinated atoms (RMSD2) means that the structural disorder is mainly confined within the second Cu solvation sphere.

The inspection of the 20 structural configurations does not reveal any water molecule organized around Cu at distances larger than 2.5 Å. This is the reason why we have defined the Cu first-coordination sphere as the set of water molecules within 2.5 Å from Cu.

It is observed that after relaxation the $M_{sqp}^0$ structures end up with an average of five water molecules in the first coordination shell, while in the $M_{oct}^0$ structures all the initially coordinated six water molecules remain within the Cu first coordination shell (see Table I). Two representative structures obtained after the nuclei position relaxation are shown in panels (c) and (d) of Figure 2. At this point the electronic density of each relaxed configuration is evaluated and it is used to calculate the corresponding theoretical XANES spectrum.

Figure 3 summarizes our results. In panel (a) we plot the experimental XANES data of $Cu^{2+}$ in water (green curve) recently taken by our group at the ESRF BM30B beamline. Together with the measured points, we display the theoretical spectra we get for the $M_{sqp}^{rel}$ configurations (red curve) and $M_{oct}^{rel}$ configurations (blue curve). The theoretical curves are obtained by averaging over the theoretical XANES spectra of the 20 configurations of each model, generated and prepared as explained above. In panel (b) we plot the derivative of the three curves.

The general agreement between experiments and theory is rather good, especially in view of the fact that no fitting of structural parameters of models $M_{sqp}^{rel}$ and $M_{oct}^{rel}$ was performed to possibly improve agreement with experimental data.

A careful inspection of the data shows that the theoretical XANES spectra computed starting from the $M_{sqp}^{rel}$ geometry are in a slightly better agreement with data than those coming from the $M_{oct}^{rel}$ geometry. This is quantitatively confirmed by the evaluation of the R quality factor that is defined as

$$R_{A-B} = \frac{\sum_{i=1}^{N} |\mu^A(E_i) - \mu^B(E_i)|}{\sum_{i=1}^{N} |\mu^A(E_i)|}, \qquad (1)$$

where $\mu^A$ and $\mu^B$ are the data of the two spectra we want to compare. We get, in fact, $R_{exp\text{-}sqp} = 7\%$ and $R_{exp\text{-}oct} = 12\%$.

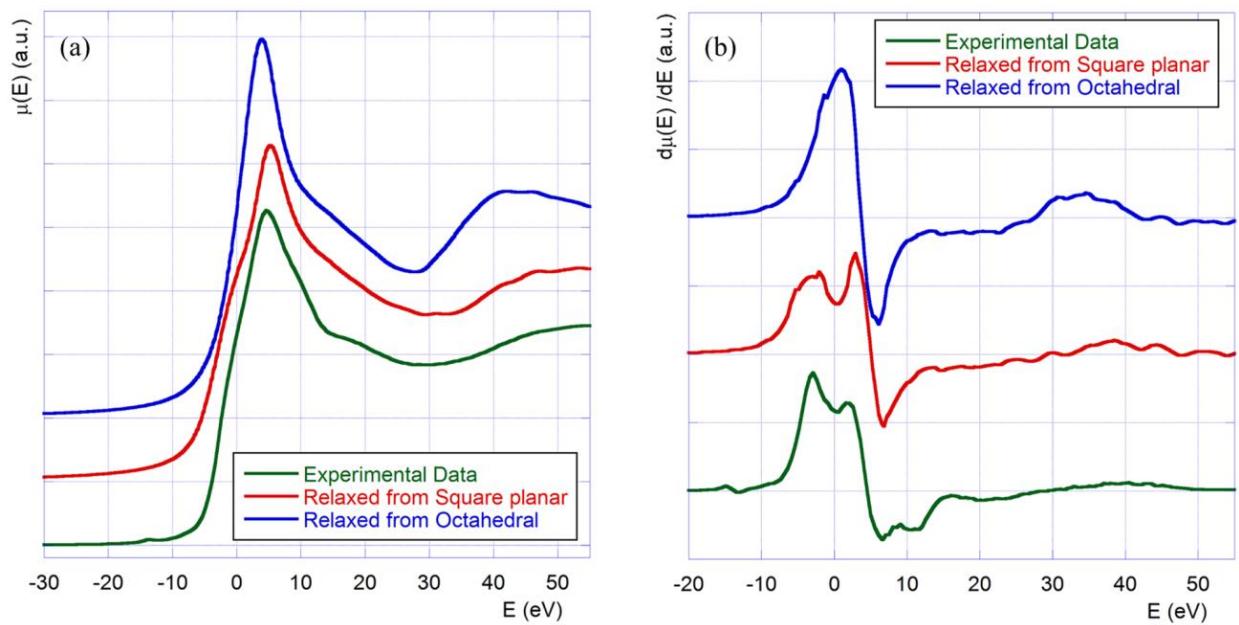

**FIG. 3**. In panel (a) the experimental data of $Cu^{2+}$ in water (green curves) are compared to theoretical results. The red curve is the average over the XANES spectra of 20 $M_{sqp}^{rel}$ configurations (after nuclei position relaxation). The blue curve is the average over the XANES spectra of 20 $M_{oct}^{rel}$ geometry configurations (after nuclei position relaxation). In panel (b) the derivative spectra are plotted. In both panels spectra are shifted along the y-axis to ease visualization.

Following the general philosophy of this paper, according to which the measured signal is the statistical average of the signals coming from the single inequivalent configurations present in the sample, we need to assess the possible simultaneous presence of the two different kinds (*sqp* and *oct*) of structures and their relative weight. To this end we construct the theoretical signal as a linear combination of the theoretical XANES spectra, separately calculated from configurations coming from the relaxed $M_{sqp}^{rel}$ geometries and those coming from the relaxed $M_{oct}^{rel}$ geometries. Taking linear combination of spectral signals is a standard procedure in XANES data analysis when more than a single structure is present in the sample (see for instance [51,52]) and gives a good insight on the relative composition. The linear combination fit is performed in the region between

-20 eV before the edge and 55 eV after the edge. The result of the analysis is shown in Figure 4 where the best fit combination is compared to the experimental data.

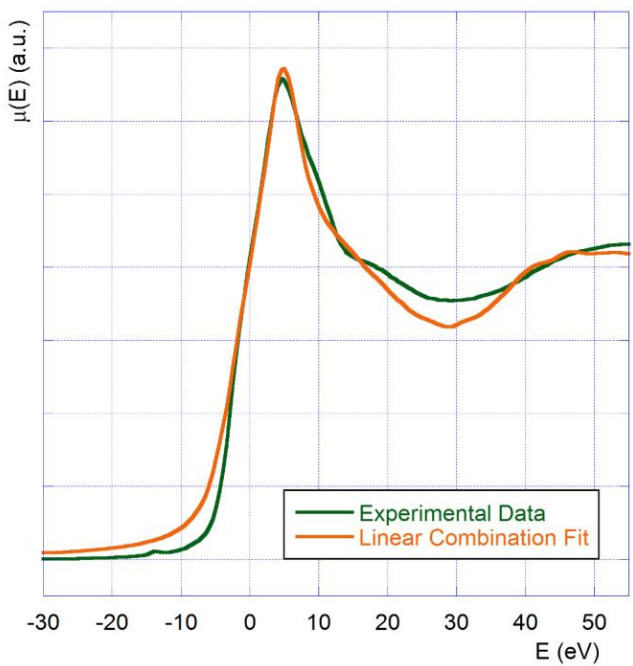

**FIG 4**. Comparison between the experimental data of $Cu^{2+}$ in water solution (green curve) and the best fit linear combination (orange curve).

The best fit is obtained by taking 73% of relaxed $M_{sqp}^{rel}$ structures and 27% of relaxed $M_{oct}^{rel}$ structures. The quality of the fit is pretty good, as confirmed by the value of the R-factor = 6%. This finding that obviously means that the majority of $Cu^{2+}$ is coordinated as in the structures coming from the "*sqp*" geometry, is well in line with the known $Cu^{2+}$ propensity for a five-fold coordination [26-28,31,53,54].

It is of interest to investigate the role of the fifth axial coordinated water in reproducing specific features of experimental data. This analysis can be done by comparing the XANES spectrum of the following three model systems: 1) a relaxed *sqp* system, $M_{29}$, including all the 29 water molecules, 2) the system, $M_5$, obtained from it by only keeping the 5 water molecules belonging to the first Cu coordination shell (conventionally defined as the set of water molecules with a Cu-O distance less than 2.5 Å) and 3) the system, $M_4$, with only 4 water molecules, in which the axial water located at a somewhat larger distance from Cu than the other four is eliminated.

Without any further adjustment of the atomic coordinates the XANES spectra of the three model systems are computed. The comparison among the three spectra is reported in Fig. 5. It can be observed that the presence of the second coordination sphere does not significantly affect the form of the spectrum (compare black and purple curve). On the contrary, in the absence of the axial water molecule the shape of the edge peak is significantly modified. It is important to notice that the presence of the second coordination sphere during the relaxation is essential to make the hydrogen bond network effective in orienting the Cu-bound water molecules in the first coordination shell. In other words, although this more distant structural network does not significantly affect the features of the XANES spectrum, it plays a crucial role in keeping in place the structure of the first coordination sphere. On the contrary, as expected, the axial water molecule has a significant impact on the electronic structure around the absorber.

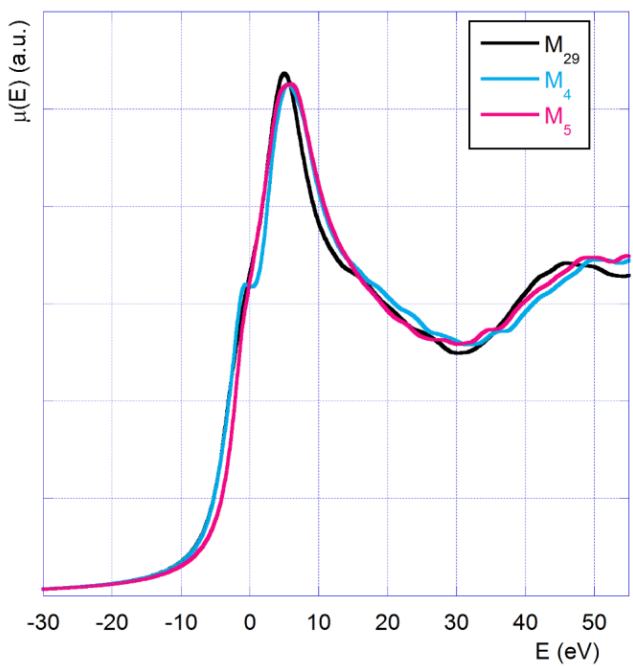

**FIG. 5** Comparison among the XANES spectra of the $M_{sqp}^{rel}$, $M_5$ and $M_4$ model systems with 29 (black curve), 5 (purple curve) and 4 (blue curve) Cu coordinated water molecules (see text for their precise definition).

The present analysis agrees with recent studies focused on $Cu^{2+}$ ion in water [26-28,31,53,54] that also suggested a predominant five-fold coordination of $Cu^{2+}$ by water molecules, excluding a significant contribution in water solution from a four-fold coordination in the ideal square-planar geometry, but not that of an axially distorted six-fold coordination in distorted octahedral geometries.

## IV. CONCLUSIONS

We have developed a general strategy aimed at providing an effective tool suited for a first-principle computation of the XANES spectrum of non-crystalline systems. The theoretical spectrum is obtained as the statistical average over a set of *ab initio* spectra each calculated using the XSpectra code of the QuantumESPRESSO suite. The key virtue of XSpectra is that it implements a DFT algorithm based on the use of ultrasoft pseudopotentials. The system configurations forming the desired statistical ensemble, for which the XANES spectrum is computed, are generated by classical MD simulations and successively subjected to a quantum-mechanical minimization step that helps relaxing unwanted local stresses.

We validate the method by calculating the K-edge XANES spectrum of $Cu^{2+}$ ions in aqueous solution showing that it compares very well with the available experimental data.

The success of our analysis suggests that a combination of MD simulations, QM minimization and DFT calculations can not only be used to calculate the XANES spectrum of transition metals in solution, but that it can also be profitably exploited in the much more difficult case of bio-molecules in complex with metals ions. In this case a geometry optimization is not affordable neither meaningful. On the other hand, a short QM relaxation is possible and has the advantage of relaxing local distortions that prevent a direct comparison between the calculated atomic structure and experimental results sensitive to local structural details. If desired, our method can be extended to employ quantum-mechanical MD simulations even in the construction of the relevant initial sample configurations. This kind of *ab initio* calculations would allow a more accurate description of the local environment around metallic centers of proteins and peptides.

## ACKNOWLEDGEMENTS

We acknowledge SUMA-INFN (Italy) for financial support.
Simulations have been performed on the INFN Zefiro cluster in Pisa (Italy).
We thank Dr. Olivier Proux for his precious support in experimental data acquisition.